\documentclass[prb,floats,twocolumn]{revtex4}

\usepackage{graphicx}
\usepackage{amsfonts}    
\usepackage{amssymb}     
\usepackage{amsbsy}      
\usepackage{amsmath}

\newcommand{\bq}{{\bf q}}

\begin{document}

\title{Correlated partial disorder in a weakly frustrated quantum antiferromagnet}

\author{M. G. Gonzalez, F. T. Lisandrini, G. G. Blesio, A. E. Trumper, C. J. Gazza, and L. O. Manuel}

\affiliation {Instituto de F\'{\i}sica Rosario (CONICET) and Universidad Nacional de Rosario,
Boulevard 27 de Febrero 210 bis, (2000) Rosario, Argentina} 

\begin{abstract} 
Partial disorder --the microscopic coexistence of long-range magnetic order and disorder-- is a rare phenomenon, 
that has been experimental and theoretically reported in some Ising- or easy plane-spin systems, driven by 
entropic effects at finite temperatures. 
Here, we present an analytical and numerical analysis of the $S=1/2$ Heisenberg antiferromagnet on the 
$\sqrt{3}\times \sqrt{3}$-distorted triangular lattice, which shows that its quantum ground state has partial 
disorder in the weakly frustrated regime. This state has a 180$^\circ$ N\'eel ordered honeycomb subsystem, 
coexisting with disordered spins at the hexagon center sites. These central spins are ferromagnetically aligned 
at short distances, as a consequence of a Casimir-like effect originated by the zero-point quantum fluctuations 
of the honeycomb lattice. 
\end{abstract}

\maketitle 

\textit{Introduction- } 
Zero-point quantum fluctuations in condensed systems are responsible for a wide variety of interesting 
phenomena, ranging from the existence of liquid helium near zero temperature to magnetically disordered Mott 
insulators \cite{balents10,zhou17}. It is in the quantum magnetism arena, precisely, 
where a plethora of control factors are available for tuning the amount of quantum fluctuations. Among these 
factors, space dimensionality, lattice coordination number, spin value $S$, and frustrating exchange 
interactions are the most relevant \cite{misguich05,lacroix11}. 

While folk wisdom visualizes zero-point quantum fluctuations like a uniform {\it foam} resulting from an 
almost random sum of states, in some cases these fluctuations contribute to the existence of very unique 
phenomena. These phenomena include semiclassical orders \cite{misguich05}, 
order by disorder \cite{henley89}, effective dimensionality reduction \cite{hayashi07,gonzalez17}, 
and topological orders associated with quantum spin liquid states \cite{balents10}, among others. 
Another role for quantum fluctuations is to allow the emergence of complex degrees of freedom from the original 
spins, like weakly coupled clusters or active spin sublattices decoupled from orphan spins. The latter has been 
proposed to explain the spin liquid behavior of the LiZn$_2$Mo$_3$O$_8$ \cite{flint13}. Here the system is 
described by a triangular spin-$\frac{1}{2}$ Heisenberg antiferromagnet which is deformed into an emergent 
honeycomb lattice weakly coupled to the central spins. 

Besides spin liquids, the presence of weakly coupled magnetic subsystems can lead to partial disorder, that is, 
the microscopic coexistence of long-range magnetic order and disorder. This rare phenomenon has been experimental 
and theoretically reported in different localized or itinerant Ising- or XY-spin highly frustrated 
systems \cite{mekata77, azaria87, diep91, plumer91, quartu97,  santamaria98, 
zheng05,zheng06,nishiwaki08,agrestini08,chern08,nishiwaki11, chen11,tomiyasu12,ishizuka12, ishizuka13,javanparast15}
and it is driven by {\it entropic effects} at finite temperature. In general, it is believed that some amount of 
spin anisotropy is needed to get partial disorder, and that the disordered subsystem behaves as a perfect paramagnet, 
with its decoupled spins justifying then the calificative of orphan spins.

\begin{figure}[ht]
\begin{center}
\includegraphics*[width=0.40\textwidth]{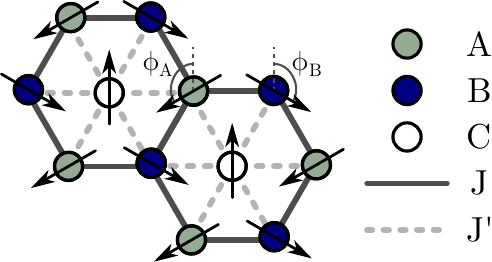}
\caption{(color online)$\sqrt{3}\times\sqrt{3}$-distorted triangular lattice, with two different exchange interactions $J$ and $J'$.
The arrows correspond to the spin directions of the semiclassical magnetic order.}
\label{fig1}
\end{center}
\end{figure}

In this work, we present an isotropic frustrated magnetic system whose ground state exhibits
partial disorder, originated by {\it zero-point quantum effects}, in contrast to the entropic origin of the so-far known cases.
Specifically, we compute the ground state of the $S=1/2$ 
antiferromagnetic Heisenberg model in the $\sqrt{3}\times \sqrt{3}$-distorted triangular lattice (Fig. \ref{fig1}), 
by means of the linear spin wave theory (LSWT) and the numerically exact density matrix renormalization 
group (DMRG). For the weakly frustrated $0\le J^{\prime}/J \lesssim 0.18$ range, we find a novel partial disorder state, 
without semiclassical analog, that consists in the coexistence of a N\'eel order in the honeycomb sublattice and 
disordered central spins. In addition, the spins of the disordered sublattice are ferromagnetically aligned at short distances, 
a correlated behavior induced, as we will show,  by a Casimir-like effect due to the ``vacuum'' 
quantum fluctuations inherent in the quantum N\'eel order of the honeycomb lattice. 

\begin{figure}[ht]
\begin{center}
\includegraphics*[width=0.30\textwidth]{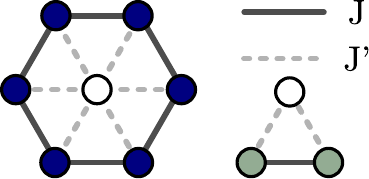}
\caption{Hexagon and triangle toy models with one spin connected to the remaining ones by a 
different exchange interaction $J'$ (dashed lines).}
\label{fig2}
\end{center}
\end{figure}

At the heart of the decoupling mechanism is the competition between the exchange energy favored
by larger coordination numbers and the zero-point quantum fluctuations. This can be 
roughly illustrated by simple toy models. For example, we can resort to the triangle and the 
hexagon with a central spin of Fig. \ref{fig2}, that present a ground state energy level 
crossing at $J'/J=1$ and $J'/J=0.63$, respectively. For small $J'$, 
the strongly connected spins form singlets, leaving the central spin completely decoupled. 

\textit{Model and methods-} 
We study the $S=\frac{1}{2}$ Heisenberg model on the $\sqrt{3}\times \sqrt{3}$-distorted triangular lattice. 
Under this distortion, the original triangular lattice is split into a honeycomb subsystem and a sublattice 
of spins at the center of each hexagon (see Fig. \ref{fig1}). Naturally,  two different nearest neighbor exchange 
interactions arise, and the Heisenberg Hamiltonian turns out
\begin{equation}
 H = J \sum_{\langle i j \rangle} {\bf S}_i \cdot {\bf S}_j + J' \sum_{\left[ i k \right]} 
 {\bf S}_i \cdot {\bf S}_k,
\label{hamiltonian}
\end{equation}
where $\langle ij\rangle$ runs over nearest neighbor spins belonging to the honeycomb lattice (with equivalent sublattices $A$ and $B$), 
while $\left[i k\right]$ links the honeycomb and central spins $C$, that interact with energy $J'.$ Throughout this work, 
we take $J = 1$ as the energy unit, while $J'$ is the only varying parameter, which we consider in the range $[0,1].$  

Almost two decades ago, this model was proposed in the context of the honeycomb reconstruction of the metallic surface of 
Pb/Ge(111) \cite{rodriguez99}. On the other hand, the complementary range $J' \ge 1$ has been considered \cite{nakano17,shimada18}, and very recently the uniform magnetization in the regime $J' \le 1$ has been computed by 
exact diagonalization \cite{shimada18b}.

Related stacked triangular lattice XY-antiferromagnets have been extensively studied in the context of the magnetic 
properties of hexagonal ABX$_3$ compounds with space group $P6_3cm$ \cite{plumer91,collins97,nishiwaki08}.

This model has two very well known limits: (i) for $J'=1,$ we recover the Heisenberg model on the isotropic triangular 
lattice, with its three equivalent sublattices and a 120$^\circ$ N\'eel ordered ground state \cite{capriotti99,white07}; while (ii) 
for $J' = 0,$ we have a honeycomb Heisenberg model with its 180$^\circ$ N\'eel ordered ground state \cite{oitmaa92}, 
and orphan (completely decoupled) spins at the centers of the hexagons.

The classical ground state of (\ref{hamiltonian}) is a simple three-sublattice order \cite{sm}, as depicted in 
Fig. \ref{fig1}, characterized by the magnetic wave vector ${\bf Q}={\bf 0}$ and by the angles 
$\phi_A = -\phi_B = - \arccos(-J'/2)$ that the spin directions on sublattices $A$ and $B$ make with the spin 
direction in sublattice $C$. This ground state evolves continuously from the honeycomb (plus orphan $C$ spins) 
to the isotropic triangular classical ground states, and it is a ferrimagnet for $0 < J' < 1$.
The Lacorre parameter \cite{lacorre87}, whose departure from the unity quantifies the degree of magnetic 
frustration, is $(2+J'^2)/(2+4J')$, so the maximal frustrated case corresponds to the isotropic triangular lattice. 

In this work, we solve (\ref{hamiltonian}) by means of complementary analytical and numerical techniques 
--the semiclassical linear spin wave theory \cite{sm} and the density matrix renormalization 
group \cite{white92}--, in order to highlight the quantum behavior without classical counterpart of the model. 
The DMRG calculations were performed on ladders of dimension $L_x \times L_y$ \cite{weichselbaum11}, 
with $L_y=6$ and $L_x$ up to 15, imposing cylindrical boundary conditions (periodic along the $y$ direction). 
We use up to 3000 DMRG states \cite{sm} in the most unfavorable case to ensure a truncation error below $10^{-6}$ in our results.

\begin{figure}[ht]
\begin{center}
\includegraphics*[width=0.40\textwidth]{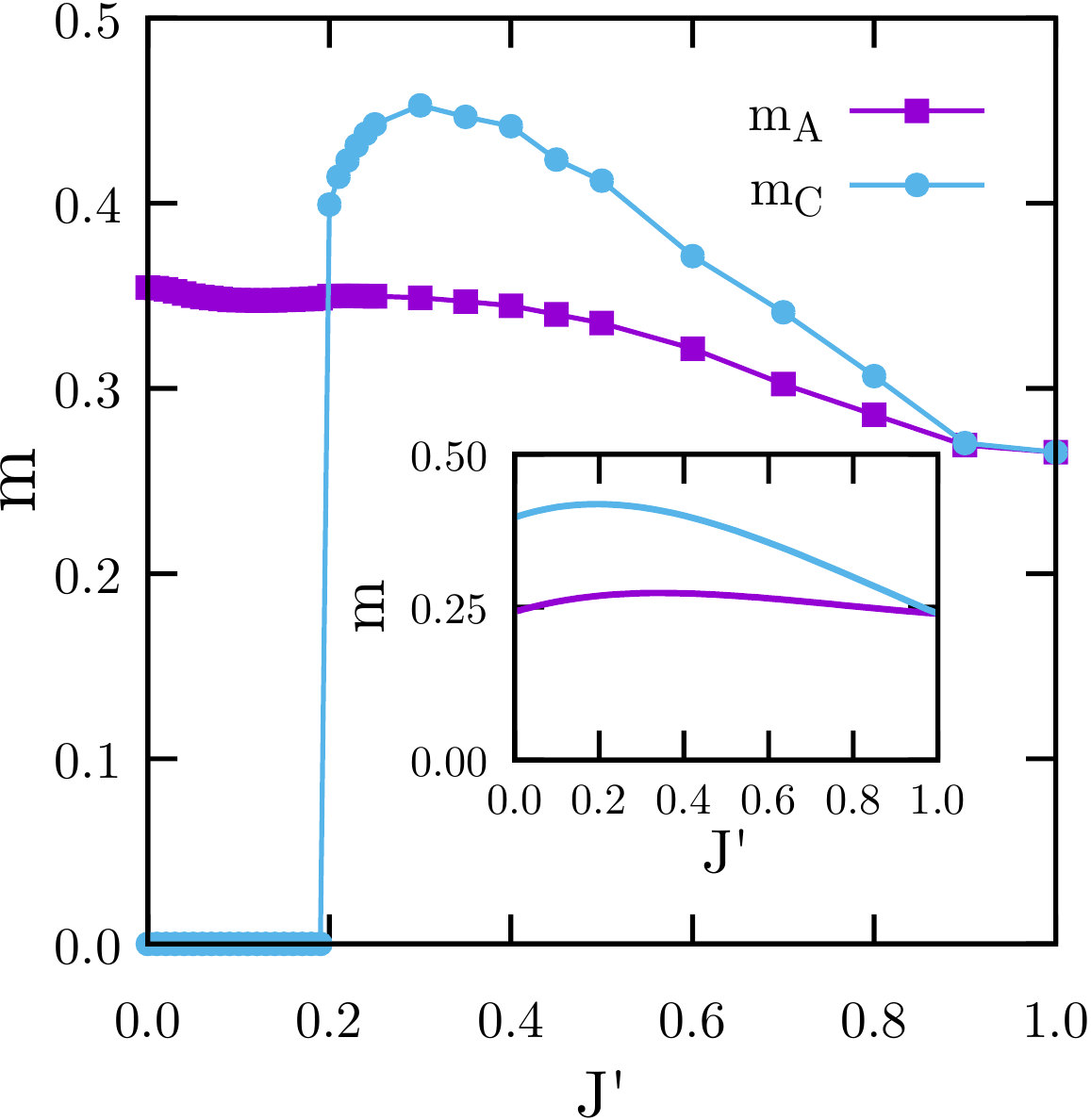}
\caption{(color online) Local magnetization of sublattices $A$ and $C$ as a function of $J',$ calculated with DMRG 
(main panel, where lines are merely a guide to the eye) and with LSWT (inset). The DMRG results correspond to the $L_y = 6,$ 
$L_x = 12$ cluster.}
\label{fig3}
\end{center}
\end{figure}

\textit{Linear spin wave results-} 
LSWT yields the same ground state magnetic structure as the classical one (see Fig. \ref{fig1}), 
with a semiclassically renormalized local magnetization $m_\alpha$ ($\alpha = A, B, C$) for each sublattice \cite{sm}, 
displayed in the inset of Fig. \ref{fig3}. As the $A$ and $B$ sublattices are equivalent, 
their order parameters coincide, while they are different from the central spin local magnetization $m_C.$ 
For $J'=0$, the central spins are decoupled from the honeycomb lattice and, consequently, $m_C$ can take any value 
from 0 to $1/2.$  As soon as $J'$ is turned on, the sublattice $C$ takes a large magnetization value, more than 
$80\%$ of its classical value, while $m_A$ varies continuously. 

Another interesting feature that can be seen is that the increase of the frustrating interaction $J'$ leads to 
an enhancement of the local magnetization in the honeycomb lattice, up to a broad maximum around $J'=0.35$ (see the darker 
curve in the inset of Fig. \ref{fig3}). 
This (apparent) paradoxical result can be explained by the increase of the effective coordination number induced by $J',$ 
that drives the system closer to its classical behavior. 
Alternatively, it can be thought that, as $J'$ is turned on, the honeycomb spins feel the $C$ subsystem as an 
uniform Weiss magnetic field $B = m_C J'$ that, through the suppression of quantum fluctuations, contributes to the 
increase of the local magnetization $m_A$ ($=m_B$), as it was found in other frustrated systems \cite{schmidt13}.
It is worth to notice that, for any $J'$, the larger order parameter belong to the sublattice $C$, which can be 
considered to be the sublattice with the smaller effective coordination number, 
$z^C_{eff} \approx 6 J'/J \le z^{A}_{eff} \approx 3 + 3 J'/J.$ This is in agreement with the fact that in 
lattices with inequivalent sites or bonds, the order parameter is lower in the sites with larger coordination 
numbers \cite{jagannathan06}. 

{\it DMRG results-}  
For all the considered range $0 \le J' \le 1$, the computed spin correlations $\left<{\bf S}_i \cdot {\bf S}_j\right>$ 
exhibit a three sublattice pattern, in full agreement with the semiclassical approach. Thus, if a given sublattice is ordered, 
all its spins will point out in the same direction (ferromagnetic order) and its local magnetization can be evaluated using the 
expression \cite{nakano13}
\begin{equation}
m_{\alpha}^2 = \frac{1}{N_\alpha(N_\alpha - 1)} \sum_{\substack{i,j\in \alpha \\ i \neq j}} 
\left< {\bf S}_i\cdot {\bf S}_j \right>, 
\label{magnetization}
\end{equation}
where $\alpha$ denotes the sublattice (A, B, or C), and $N_\alpha$ is its number of sites. 
The calculated $m_A$ and $m_C$ are shown in the main panel of Fig. \ref{fig3} for the $N_s=12\times 6$ cluster (for other cluster sizes, we have obtained a similar result \cite{sm}).

The most eye-catching difference between the DMRG and the semiclassical local magnetizations appears in the 
weakly frustrated parameter region, close to the honeycomb phase, $ 0 \le J' \lesssim 0.18$. 
There, DMRG shows a vanishing order parameter $m_C$ for the $C$ sublattice along with an almost constant honeycomb lattice 
local magnetization $m_A$. This corresponds to a partially disordered phase, driven solely by quantum fluctuations (in 
competition with frustration), as we are working at $T=0.$ Notice that, in general, partially disordered phases are 
associated with entropic effects, and they appear at intermediate temperatures, between the lower and higher energy scales of 
the system \cite{diep91,nishiwaki11,ishizuka12}. 

There is a critical value $J'_c \simeq 0.18$ where the $C$ sublattice gets suddenly ordered, as it happens at $J'=0^+$ 
in LSWT (inset of Fig. \ref{fig3}). Furthermore, beyond this critical value, the DMRG calculations show a higher 
local magnetization in the central spin sublattice, in agreement with the semiclassical expectation \cite{jagannathan06}.

The local magnetization $m_A$ ($ = m_B$) of the honeycomb spins decreases when $J'$ increases due to the frustration 
introduced by the coupling with the central spins, until it reaches its minimum value in the isotropic triangular lattice, 
corresponding to the most frustrated case \cite{magnet}. In contrast with the spin wave results, $m_A$ does not exhibit 
a clear maximum for intermediate values of $J'$, but shows an almost constant region ranging from $J'=0$ to $0.2$. 
This feature signals a negligible effect of the central spins on the honeycomb ones.

As we have mentioned above, the semiclassical ground state is ferrimagnetic for $0 < J' < 1.$ So, in order to further 
characterize the DMRG ground state magnetic structure, we calculate the lowest eigenenergy in the different $S_z$ subspaces. 
In the case of $J' = 0$, the spins $C$ are totally disconnected from the honeycomb subsystem and, thus, do not contribute 
to the total energy. This results in a perfect paramagnetic behavior of the central spin sublattice, with a high degeneracy of the ground 
state, as $E(S_z=0) = E(S_z=\pm 1) = ... =E(S_z=S_z^{max}=\pm 1/2\times N_s/3),$ where $S_z^{max}$ denotes the maximum value of 
$S_z$ whose subspace belongs to the ground state manifold, and $N_s$ is the number of sites of the 
cluster. For $J' \neq 0$ the situation changes: for $J' \le J'_c$ there are no more orphan spins and $S_z^{max}=0$ (see Fig. S2(a)~\cite{sm}), indicating that the partially disordered phase is a (correlated) singlet. On the other hand, at the critical value $J'_c,$ 
$S_z^{max}$ jumps to a finite value, signaling a first-order transition from the singlet to the ferrimagnetic state (see Fig. S2(b)~\cite{sm}).  
With further increase of $J'$, $S_z^{max}$ decreases until it vanishes for the isotropic 
triangular point $J'=1,$ whose ground state again is a singlet. Therefore, the DMRG magnetic order has a ferrimagnetic 
character for $J'_c < J' < 1.$ This behavior shows little finite size effects~\cite{sm} and it quantitatively agrees with the exact diagonalization predictions \cite{shimada18b}.

Next, we calculate the DMRG angles between the local magnetization in different sublattices. 
Previously, we have checked that the classical angles $\phi_\alpha$ are not renormalized in LSWT, in contrast to what 
happens in other frustrated models, like the triangular anisotropic Heisenberg model, where the pitch of its spiral 
magnetic order is sizable renormalized by quantum fluctuations \cite{manuel99}. 
In order to estimate the angles, we take into account that the large values of the DMRG local magnetizations enable us to 
use a semiclassical picture of the spins. Let us think of a three-spin unit cell built by spins $A$, $B,$ and $C$ of lengths 
$m_A$, $m_B,$ and $m_C$, respectively, as seen in Fig. \ref{fig1}.
We can assume that the $S_z^{max}$ subspace corresponds to the $C$ spin pointing out in the $z$-direction, while the $A$ and 
$B$ spins make angles $\phi_A = -\phi_B$ with it. Then, we get the equation
$$\frac{N_s}{3} \left( m_C + 2 m_A  \cos \phi_B\right)  = S_z^{max},$$
for $\phi_B$ \cite{rodriguez99}, which finally leads to the angle $\theta$ between $A$ and $B$ spins
\begin{equation}
\theta = 2 \arccos \left[ \frac{1}{2 m_A} \left( m_C - \frac{3S_z^{max}}{N_s} \right)\right].
\label{angleres}
\end{equation}
\begin{figure}[ht]
\begin{center}
\includegraphics*[width=0.40\textwidth]{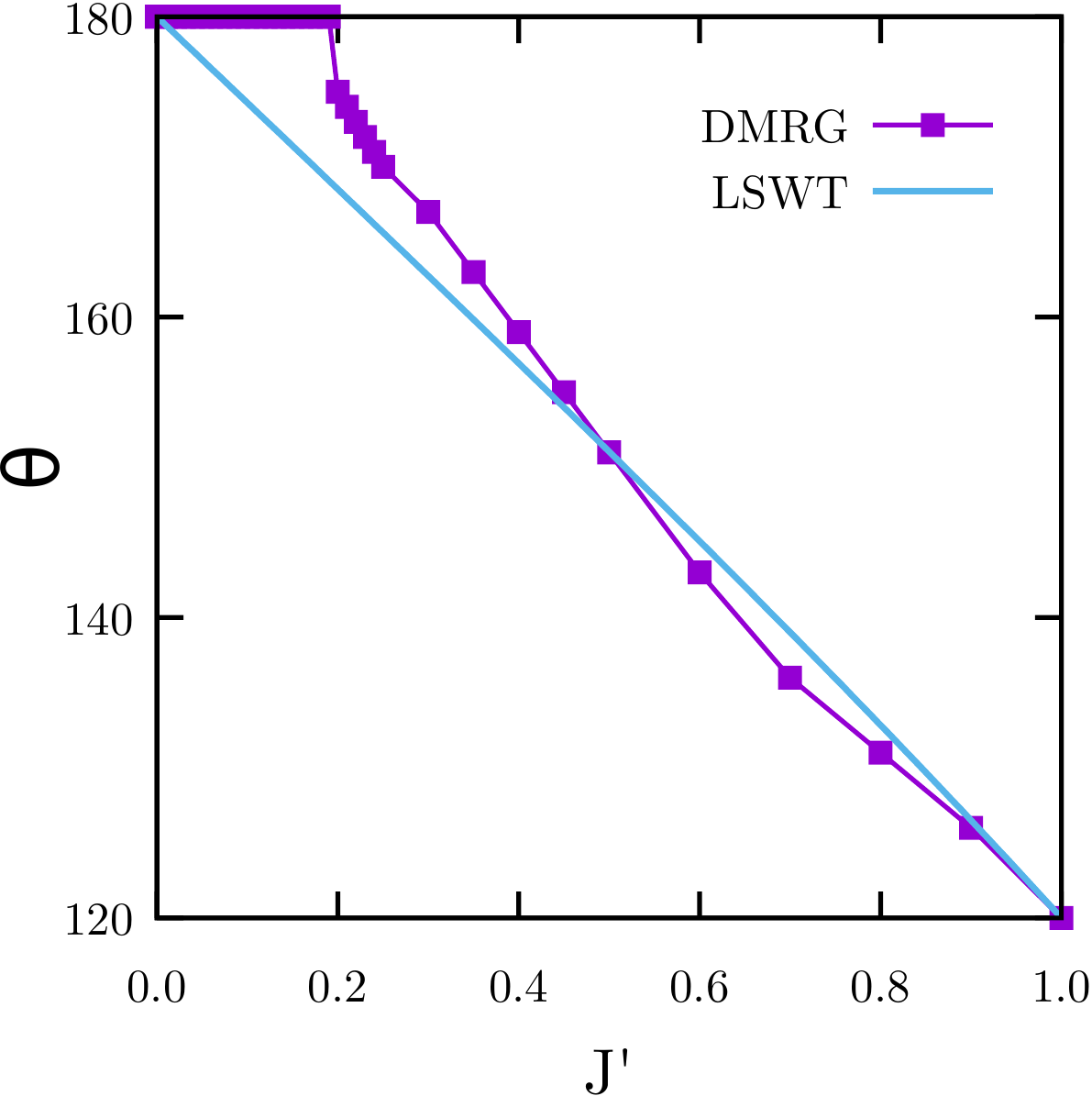}
\caption{(color online) DMRG and LSWT angles $\theta$ between spins in the $A$ and $B$ sublattices as a function of $J'$.}
\label{fig4}
\end{center}
\end{figure} 

In Fig. \ref{fig4} the angle $\theta$ is plot as a function of $J'.$ It can be seen that the honeycomb 180$^\circ$ N\'eel 
order persists all along the partially disordered phase where $m_c$ vanishes. This result is a simple consequence of the 
singlet character of the ground state for $0 < J' < J'_c$ ($S_z^{max}=0$ in Eq. \ref{angleres}). As a consequence, 
the canting behavior observed semiclassically for any finite $J'$ moves to the region above $J'_c$ in the strong quantum 
limit $S=1/2$ calculated with DMRG. That is to say that, when $J'$ is small, $C$ spins are disordered because the system gains 
zero-point quantum energy from that disorder. For larger values of $J'$, the system chooses to gain 
(frustrated) exchange energy over zero-point quantum fluctuation, and the $C$ sublattice gets ordered, canting simultaneously 
the $A$ and $B$ spins. 

It is worth to emphasize that, even when the ground state seems to undergo a first-order transition at $J'_c$ 
($S_z^{max}$ changes abruptly and the local magnetization $m_C$ sharply rises), the angle $\theta$ between the $A$ and $B$ 
spins varies continuously from its $180^\circ$ value in the partially disordered phase. This is similar to the spin wave 
behavior around $J'=0,$ where the sublattice $C$ is disordered, but as soon as $J'$ rises, $m_C$ suddenly grows over $m_A$ 
without any abrupt change in the magnetic order. 

The quantitative agreement between the DMRG and LSWT angles for $J' \gtrsim J'_c,$ displayed in Fig. \ref{fig4}, is a clear 
evidence that, beyond the partially disordered phase present in the weakly frustrated regime, the quantum ground state of the 
model is very well described semiclassically. 

\begin{figure}[ht]
\begin{center}
\includegraphics*[width=0.40\textwidth]{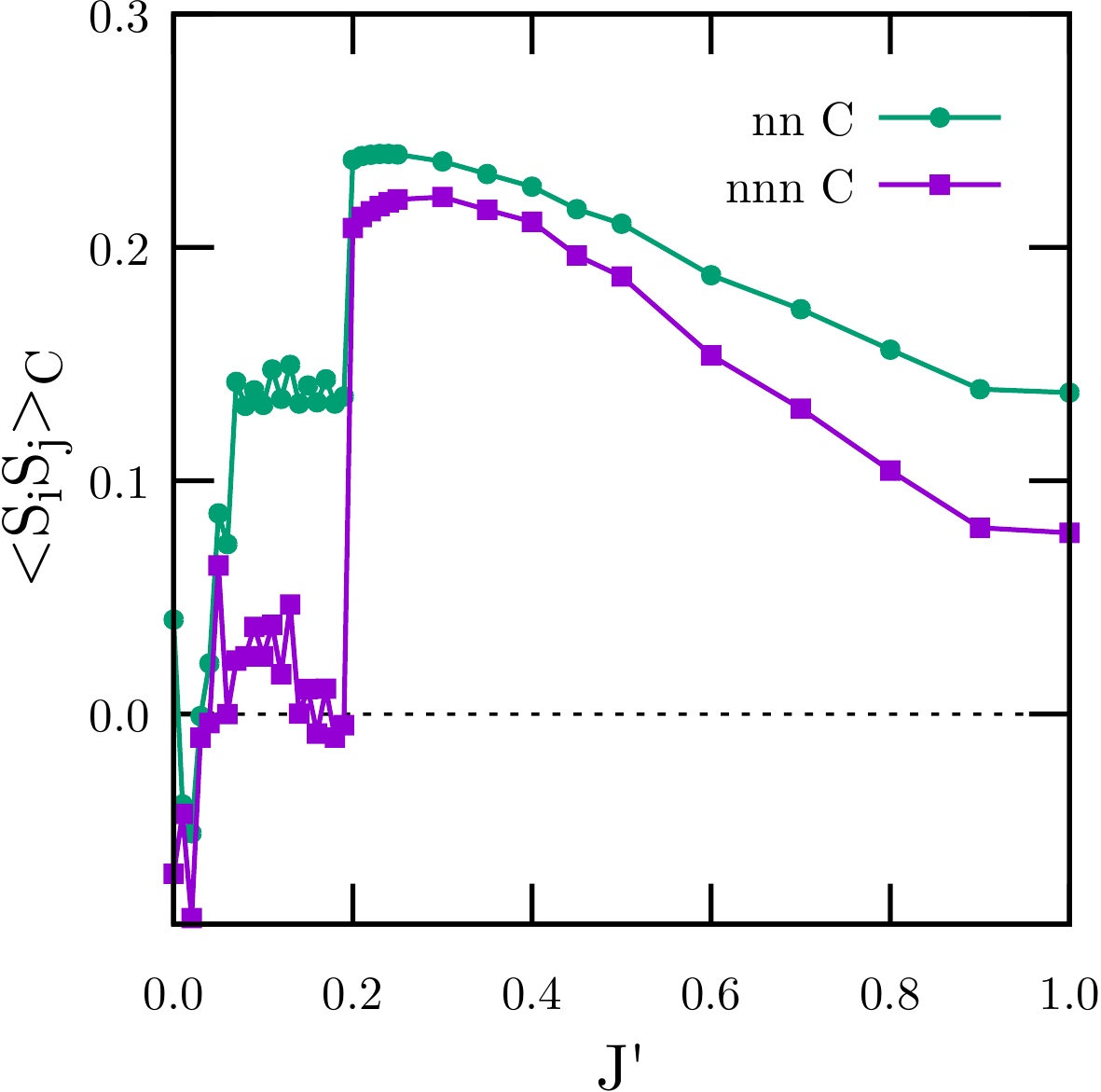}
\caption{Average DMRG nearest neighbor (nn) and next-nearest neighbor (nnn) correlations between central spins as a function of $J'$.}
\label{fig5}
\end{center}
\end{figure}

Up to now, we have characterized the region between $J'=0$ and $0.18$ as a singlet partially disordered phase. To deepen the 
understanding of such disorder, in Fig. \ref{fig5} the average nearest (nn) 
and next-nearest (nnn) neighbor spin correlations \cite{average} between the 
central spins is shown as  a function of $J'.$ 
It can be seen that, even though the sum over all the correlations in the $C$ sublattice is zero in the region of its null local magnetization (see Eq. \ref{magnetization}), the nn 
correlation has an almost constant positive value, close to 1/8, while the nnn correlation is close to zero. This means that central spins exhibit (very) short range ferromagnetic correlations between them, suggesting that the partial disordered phase may be thought as a sort of a {\it resonating} spin-triplet valence bond state \cite{shindou09}.
In other magnetic systems which exhibit partial disorder, mostly with Ising or XY-like spins, the disordered subsystem is a 
perfect paramagnet of orphan spins, with zero correlation between them  \cite{diep91,plumer91}. 
As there is no explicit exchange interaction between the $C$ spins, their correlation should be mediated by the 
coexistent honeycomb N\'eel order.

In order to build up a qualitative argument about the origin of the correlated character of the partially disordered phase 
below $J'_c$, we appeal to a Weiss molecular field approach for the simplest toy model (see Supplemental Material \cite{sm} 
for details). 
We consider a 4-spin cluster,  composed of two nearest neighbor honeycomb spins (1 and 2) interacting with the two closest $C$ 
spins (3 and 4), with a Hamiltonian 
$$ H= J{\bf S}_1\cdot {\bf S}_2 + J'\left({\bf S}_1+{\bf S}_2\right)\cdot\left({\bf S}_3+{\bf S}_4\right).$$
After fixing the state of honeycomb spins 1 and 2 as a ``classical N\'eel order'', plus zero-point 
quantum fluctuations quantified by a parameter $r$, we arrive at an effective Hamiltonian for the central spins 3 and 4, 
that consists of a Zeeman term associated with an effective uniform magnetic field $B \propto rJ',$ perpendicular to the 
honeycomb N\'eel order.  Hence, this toy model helps us to understand how the ferromagnetic correlations between 
nearest neighbor central spins are built up under an effective interaction between them, driven by the vacuum 
fluctuations of the honeycomb N\'eel order; that is, the correlation between central spins can be considered a 
Casimir-like effect. This argument is further supported by the presence of nearest-neighbor antiferromagnetic spin-spin correlations between the orphan spins and the honeycomb ones (see Ref.~\cite{sm}).
This treatment is valid whenever this N\'eel order is unaffected by the feedback of the spins 3 and 4. This seems to 
be the case in the DMRG calculations for the lattice, as the local magnetization $m_A$ changes only slightly by the 
coupling of the $C$ spins in the partially disordered phase (see Fig. \ref{fig3}). 
Also, the toy model explains the almost constant nearest neighbor correlation between central spins that can be seen, below $J'_c,$ 
in Fig. \ref{fig5} (except for the non-monotonic behavior very close to $J'=0$, probably due to numerical inaccuracies). 
It should be mentioned that, due to the singlet character of the DMRG ground state, the correlations
between the central spins are isotropic, and not perpendicular to a given direction like in the toy model.

The absence of long-range ferromagnetic order of the $C$ subsystem below $J'_c$ \cite{campo} can be
roughly explained as follows: in the previous Weiss mean-field argument, the zero-point quantum fluctuations of the honeycomb subsystem act as a magnetic field for the $C$ spins. The direction of this effective molecular magnetic field is  "random" as it depends on the phase fluctuations of the departure from the N\'eel state. 
Therefore, while the nearest-neighbor $C$ spins are ferromagnetically correlated, the overall subsystem remains disordered due to the "randomness" of the zero point quantum fluctuations.

\textit{Summary-} 
We have studied a $S=1/2$ Heisenberg antiferromagnet with inequivalent exchange interactions 
on a distorted triangular lattice which, in the weakly frustrated regime $0 \le J'/J \lesssim 0.18$, 
exhibits a novel correlated partial disordered phase, driven by the competition between zero-point 
quantum fluctuations and frustration. 
Even if partial disordered phases were already known \cite{mekata77, azaria87, diep91, plumer91, quartu97,  santamaria98, 
zheng05,zheng06,nishiwaki08,agrestini08,chern08,nishiwaki11, chen11,tomiyasu12,ishizuka12, ishizuka13,javanparast15}--in anisotropic systems at finite temperatures and 
with disordered subsystems that behave as perfect paramagnets--,   
here we uncover, for the first time, a partially disordered phase as 
the ground state of a simple isotropic Heisenberg antiferromagnet. This phase exhibits 
the coexistence of two magnetic subsystems: one antiferromagnetically ordered, and the other, disordered with 
its spins ferromagnetically correlated at short distances due to the zero-point quantum fluctuations of the ordered subsystem \emph{via} a Casimir-like effect. 

\textit{Acknowledgements-} 
We thank Dr. H. Nakano for calling our attention to Ref. \cite{shimada18b}. 
This work was supported by CONICET under grant Nro. 364 (PIP2015).


 \widetext
 \pagebreak
 \setcounter{equation}{0}
 \setcounter{figure}{0}
 \setcounter{page}{1}
 \makeatletter
 \renewcommand{\theequation}{S\arabic{equation}}
 \renewcommand{\thefigure}{S\arabic{figure}}
 \renewcommand{\bibnumfmt}[1]{[S#1]}
 \renewcommand{\citenumfont}[1]{S#1}
 \begin{center}
 {\bf \large Supplemental material for: \\Correlated partial disorder in a weakly frustrated quantum antiferromagnet}
 \end{center}
 \begin{center}
 {M. G. Gonzalez, F. T. Lisandrini, G. J. Blesio, A. E. Trumper, C. J. Gazza, and L. O. Manuel}
 \end{center}
 \begin{center}
 {\it Instituto de F\'{\i}sica Rosario (CONICET) and Universidad Nacional de Rosario,  
 Boulevard 27 de Febrero 210 bis, (2000) Rosario, Argentina}
 \end{center}
We sketch the linear spin wave calculation for the Heisenberg model on lattices with complex unit cells, as it 
corresponds to the $\sqrt{3}\times\sqrt{3}$-distorted triangular lattice. Also, we present the Weiss mean field-like 
treatment of a 4-spin cluster model in order to understand the role of quantum fluctuations of the honeycomb N\'eel 
order in the establishment of ferromagnetic correlations between central spins in the partially disordered phase. 
Finally, we present some DMRG results related to the convergence with the number of states kept, finite size effects, the determination of $S_z^{\rm max}$, and 
the spin-spin correlation between honeycomb and C spins.

\vskip 1cm

\section{Linear spin wave calculation}

In this section, we present the linear spin wave theory used to solve semiclassically the Heisenberg model. 
The Hamiltonian (1) can be rewritten in a generic way as
\begin{equation}
 H = \frac{1}{2}\sum_{i j \alpha \beta}J_{\alpha \beta}({\bf R}_i-{\bf R}_j){\bf S}_{i\alpha}\cdot {\bf S}_{j\beta},
\label{heis} 
\end{equation}
where $i, j$ denote unit cells with vector positions ${\bf R}_i$, ${\bf R}_j$, respectively, while $\alpha, \beta$ 
represent the spins inside the unit cell. $J_{\alpha \beta}({\bf R}_i-{\bf R}_j)$ is the exchange interaction
between the spin $\alpha$ in the unit cell ${\bf R}_i$ and the spin $\beta$ in the unit cell ${\bf R}_j.$
In our case, the $\sqrt{3}\times \sqrt{3}$-distorted triangular lattice is a non-Bravais lattice that can be 
described as a triangular lattice with basis vectors ${\bf a}_1 = \left(\frac{3a}{2},\frac{\sqrt{3}a}{2}\right)$ 
and ${\bf a}_2 = (0,\sqrt{3}a)$ ($a$ is the nearest neighbor distance), and a unit cell with the three sites 
A, B, and C (see Fig. 1). $J_{\alpha \beta}$'s only take non-zero values for nearest neighbors: 
$$
\begin{array}{lll}
J_{AB}({\bf 0}) = J_{AB}({\bf a}_1) = J_{AB}({\bf a}_1-{\bf a}_2)   =  J,\\
\\
J_{AC}({\bf 0})=J_{BC}({\bf 0})=J_{AC}({\bf a}_2)=J_{BC}({\bf a}_2)=J_{AC}({\bf a}_1)=J_{BC}({\bf a}_2-{\bf a}_1)  = J',
\end{array}
$$
and the remaining ones can be obtained through the relation 
$J_{\alpha \beta}({\bf R}_i-{\bf R}_j) = J_{\beta \alpha}({\bf R}_j-{\bf R}_i)$.

In order to perform the spin wave analysis, we first need the classical ground state of the Heisenberg model.
For this purpose, we consider classical spin vectors in (\ref{heis}) and we minimize the corresponding 
energy \cite{nagamiya67}
\begin{equation}
 E_{\rm clas} = \sum_{\alpha \beta {\bq}} J_{\alpha \beta}({\bq}){\bf S}_{\bq \alpha}\cdot {\bf S}_{-{\bq}\beta},
 \label{eclas}
\end{equation}
where $J_{\alpha \beta}({\bq})=\sum_{{\bf R}} J_{\alpha \beta}({\bf R})e^{i{\bq}\cdot {\bf R}}$ is the Fourier 
transform of the exchange interaction matrix $J,$ and 
${\bf S}_{\bq \alpha}=\frac{1}{\sqrt{N}}\sum_{i}{\bf S}_{i\alpha}e^{i{\bq}\cdot {\bf R}_i}$ are classical vectors 
subject to the normalization condition ${\bf S}_{ i\alpha}^2 =1$. The sum over ${\bf R}$ runs over all the $N$ 
unit cells in the cluster. 
For the $\sqrt{3}\times\sqrt{3}$-distorted triangular lattice, the minimization of the classical energy (\ref{eclas}) 
gives the magnetic wave vector ${\bf Q}={\bf 0},$ implying that the same absolute magnetic pattern is repeated in 
each unit cell. If $\phi_\alpha$ is the angle between the spins in the $\alpha$ sublattice and the $z$ direction, 
after a little algebra we find $\phi_A = -\phi_B = -\arccos (-J'/2J),$ and $\phi_C=0.$ This three sublattice (coplanar) 
structure is shown in Fig. 1. 

Once we know the classical ground state magnetic structure, we make a rotation to local axes in (\ref{heis}). 
That is, at each site we define a local frame in such a way that the classical order points along the local $z$ axis.
If, for simplicity, we consider a planar classical magnetic structure (like in our case), lying in the $xz$ plane, 
then the spin operators in the local axes are
$$
 \tilde{\bf S}_{i\alpha} = {\cal R}_{y}(\Theta_{i\alpha}){\bf S}_{i\alpha}, 
$$
where $\Theta_{i\alpha}={\bf Q}\cdot{\bf R}_i + \phi_{\alpha}$ is the angle between the classical ${\bf S}_{i\alpha}$ 
spin and the global $z$ direction, and ${\cal R}_{y}(\Theta_{i\alpha})$ is the matrix rotation of angle 
$\Theta_{i\alpha}$ around the $y$ direction. In terms of the local spin operators $\tilde{\bf S}$, the Hamiltonian 
(\ref{heis}) turns into 
$$
 H = \frac{1}{2}\sum_{i j \alpha \beta} \left[\cos(\Theta_{i\alpha}\!-\!\Theta_{j\beta}) \left( \tilde{S}^x_{i \alpha}
 \tilde{S}^x_{j \beta}+ \tilde{S}^z_{i \alpha} \tilde{S}^z_{j \beta}\right)+  \sin(\Theta_{i\alpha}\!-\!\Theta_{j\beta}) 
 \left( \tilde{S}^z_{i \alpha}  \tilde{S}^x_{j \beta}- \tilde{S}^x_{i \alpha} \tilde{S}^z_{j \beta}\right)
 +\tilde{S}^y_{i \alpha}\tilde{S}^y_{j \beta} \right].
$$

In the local axis frame, the classical magnetic structure is a ferromagnetic one along the $z$ direction. Now, we can 
represent the local spins by bosonic operators $a_{i\alpha}, a^\dagger_{i\alpha},$ using the typical Holstein-Primakoff 
transformation \cite{holstein40}:
\begin{eqnarray}
\tilde{S}^z_{i\alpha} & = & S - a^\dagger_{i\alpha}a_{i\alpha},\nonumber \\
\tilde{S}^+_{i\alpha} & = & \sqrt{2S-a^\dagger_{i\alpha}a_{i\alpha}}\;a_{i\alpha}, \label{hp}\\ 
\tilde{S}^-_{i\alpha} & = & a^\dagger_{i\alpha}\;\sqrt{2S-a^\dagger_{i\alpha}a_{i\alpha}}\nonumber.
\end{eqnarray}

In the semiclassical approximation, valid for $S \to \infty$, the (operator) square roots that appear in the 
Holstein-Primakoff representation (\ref{hp}) are replaced by the scalar $\sqrt{2S},$  yielding the linear spin wave 
approximation. After performing this approximation and a Fourier transformation to momentum space, the Hamiltonian 
(\ref{heis}) is written as
\begin{equation}
 H_{LSW} = \left(1+\frac{1}{S}\right)E_{\rm clas} + \frac{S}{2} \sum_{\alpha \beta {\bq}}
 {\bf a}^\dagger_{-{\bq}\alpha} \cdot {\cal D}_{\alpha \beta}({\bq})\cdot{\bf a}_{\bq \beta}, 
 \label{hlsw}
\end{equation}
where the spinor ${\bf a}_{\bq \beta} = (a_{{\bq}\beta}, a^\dagger_{-{\bq} \beta})^T$, the bosonic dynamical matrix is 
$$
 {\cal D}_{{\bq}} = \left(
 \begin{array}{cc}
  A_{\alpha \beta}({\bq}) & -B_{\alpha \beta}({\bq}) \\ 
 -B_{\alpha \beta}({\bq}) & A_{\alpha \beta}({\bq}) 
 \end{array}
\right),
$$
and
\begin{eqnarray*}
 A_{\alpha\beta}({\bq}) & = & \sum_{\bf R} J_{\alpha \beta}({\bf R})\cos (\bq \cdot {\bf R}) 
 \cos^2\left(\frac{{\bf Q}\cdot{\bf R}+\phi_\alpha-\phi_\beta}{2}\right)-\delta_{\alpha\beta}\sum_{\gamma \bf R}J_{\alpha \gamma}({\bf R})
 \cos \left({\bf Q}\cdot {\bf R}+ \phi_\alpha-\phi_\gamma\right), \\
 B_{\alpha \beta}(\bq) & = & \sum_{\bf R} J_{\alpha \beta}({\bf R}) \cos (\bq \cdot {\bf R}) 
 \sin^2\left(\frac{{\bf Q}\cdot{\bf R}+\phi_\alpha-\phi_\beta}{2}\right).
\end{eqnarray*}

The quadratic Hamiltonian (\ref{hlsw}) can be diagonalized by means of a paraunitary Bogoliubov transformation \cite{colpa78}.
After this procedure, the order parameter at each site can be evaluated using $m_{i\alpha} = |<\tilde{S}^z_{i\alpha}>|.$

\section{Weiss mean field-like approximation for a 4-spin cluster}

\begin{figure}[ht]
\begin{center}
\includegraphics*[width=0.10\textwidth]{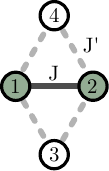}
\caption{4-spin cluster model. The solid line corresponds to an interaction $J$, while the dashed ones correspond 
to $J'$.}
\label{figS1}
\end{center}
\end{figure}

In order to understand the origin of the correlated character of the partial disordered phase below $J'_c \simeq 0.18$, 
we resort to a molecular Weiss-like mean field approach for a simple toy model. 
We consider a 4-spin cluster as it is shown in Fig. \ref{figS1}, composed of two central spins C (3 and 4) and two 
honeycomb spins, one A (1) and another B (2). Its Hamiltonian is 
\begin{equation}
 H = J {\bf S}_1 \cdot {\bf S}_2 + J'\left({\bf S}_1 + {\bf S}_2\right)\cdot \left({\bf S}_3 + {\bf S}_4\right),
 \label{h4spins}
\end{equation}
and we are interested in the case $J' \ll J$.

The exact ground state of Hamiltonian (\ref{h4spins}) can be easily found, and it consists, for $J'/J \le 1/2,$ \cite{nota}  
in a singlet state between the 1 and 2 spins, while the remaining spins 3 and 4 can have any spin projection 
$\sigma, \sigma'$:
$$
|\rm GS>=\left(\frac{|1\uparrow 2 \downarrow>-|1\downarrow 2\uparrow>}{\sqrt{2}}\right)\otimes| 3\sigma 4\sigma'>.
$$
Naturally, the average spin correlation between 3 and 4 is zero, and this model corresponds to another example, 
like the ones presented in Fig. 2, where the less connected spins are completely decoupled.
However, we are not interested in the exact solution of (\ref{h4spins}) --with its singlet state between spins 1 and 2--, 
but in an approximated treatment that mimics what happens in the partially disordered phase of the 
$\sqrt{3}\times \sqrt{3}$-distorted triangular Heisenberg model.

As DMRG numerical results indicate that the honeycomb N\'eel order remains almost unaltered for 
$ 0 \le J' \lesssim 0.18,$ we propose to fix the spins 1 and 2 in the normalized state
\begin{equation}
|\psi(1,2)> = \frac{1}{\sqrt{1+2r^2}} \left[|1\uparrow 2\downarrow> + r\left(|1 \uparrow 2\uparrow> + 
1\downarrow 2\downarrow>\right)\right],
\label{estado}
\end{equation}
that represent a ``classical N\'eel order'' (spin 1 $\uparrow$ and spin 2 $\downarrow$) plus quantum fluctuations 
quantified by the real parameter $r$. $r$ can be related to the local magnetization of the ``honeycomb spins'' as
$$
m_0 = <\psi(1,2)|S_1^z|\psi(1,2)> = \frac{1}{2(1+2r^2)},
$$
so we get the relation $r \simeq \sqrt{\frac{1}{2}-m_0} \ll 1$ for $m_0$ close to its classical value.

In the spirit of the Weiss molecular field theory, we approximate (\ref{h4spins}) by an effective Hamiltonian for 
spins 3 and 4, 
\begin{equation}
 H_{34} = J <\psi(1,2)|{\bf S}_1 \cdot {\bf S}_2|\psi(1,2)> + J'<\psi(1,2)|\left({\bf S}_1 + {\bf S}_2\right)|\psi(1,2)> 
 \cdot \left({\bf S}_3 + {\bf S}_4\right), 
\label{ham34}
\end{equation}
which results in the expression
\begin{equation}
 H_{34} = 4 m_0 r J' \left(S_3^x + S_4^x\right) + {\rm const.}
 \label{h34x}
\end{equation}
That is, the quantum fluctuations of the classical N\'eel order of the spins 1 and 2 act as an effective uniform 
magnetic field in the $x$ direction for the central spins 3 and 4. For any $r > 0$ the ground state of (\ref{h34x}) is 
$|3 \downarrow 4\downarrow>_x,$ where $x$ refers to the quantization axis. 

Of course, the $x$ and $y$ directions are equivalent, the same as the spin projection.  
To show this, we consider a more general ``N\'eel state'' than (\ref{estado}), giving the possibility that the 
zero-point quantum fluctuation states $|1\uparrow 2\uparrow>$ and $|1\downarrow 2\downarrow>$ are out of phase, that is 
\begin{equation}
 |\psi(1,2)> = \frac{1}{\sqrt{1+2r^2}} \left[|1\uparrow 2\downarrow> + r\left(e^{i\theta_\uparrow}|1 \uparrow 2\uparrow> + 
 e^{i\theta_\downarrow} |1\downarrow 2\downarrow>\right)\right].
 \label{estado2}
\end{equation}
Inserting this state in (\ref{ham34}), we get the effective Hamiltonian 
$$
 H_{34} = 2 m_0 rJ' \left[\left(\cos\theta_\uparrow+\cos\theta_\downarrow\right)\left(S_3^x+S_4^x\right) + 
 \left(\sin\theta_\downarrow-\sin\theta_\uparrow\right)\left(S_3^y + S_4^y\right)\right], 
$$
that is, a Zeeman term for spins 3 and 4 with an effective magnetic field  
$\vec{B}=  2m_0 rJ' \left(\cos\theta_\uparrow+\cos\theta_\downarrow,\sin\theta_\downarrow-\sin\theta_\uparrow,0\right),$
perpendicular to the classical N\'eel order in the $z$ direction.  
Again, in the ground state $|\psi(3,4)>$ of this Hamiltonian the spins 3 and 4 are aligned along the $\vec{B}$ direction 
and, straightforwardly, it results in the classical ferromagnetic correlation 
 $$
 <\psi(3,4)|{\bf S}_3 \cdot {\bf S}_4|\psi(3,4)>= \frac{1}{4} 
 $$
between them, independently of the value of $J'$. It is worth to notice that in the two extreme cases, when the state of 
spins 1 and 2 is taken as a classical N\'eel one or as a singlet, there is no correlation between spins 3 and 4. So, 
a {\it quantum} N\'eel order is necessary  for the spins 3 and 4 to be correlated.

\section{DMRG results}

\subsection{Ground state energy as a function of $S_z$: determination of $S_z^{\rm max}$}

\begin{figure}[ht]
\begin{center}
\includegraphics*[width=0.70\textwidth]{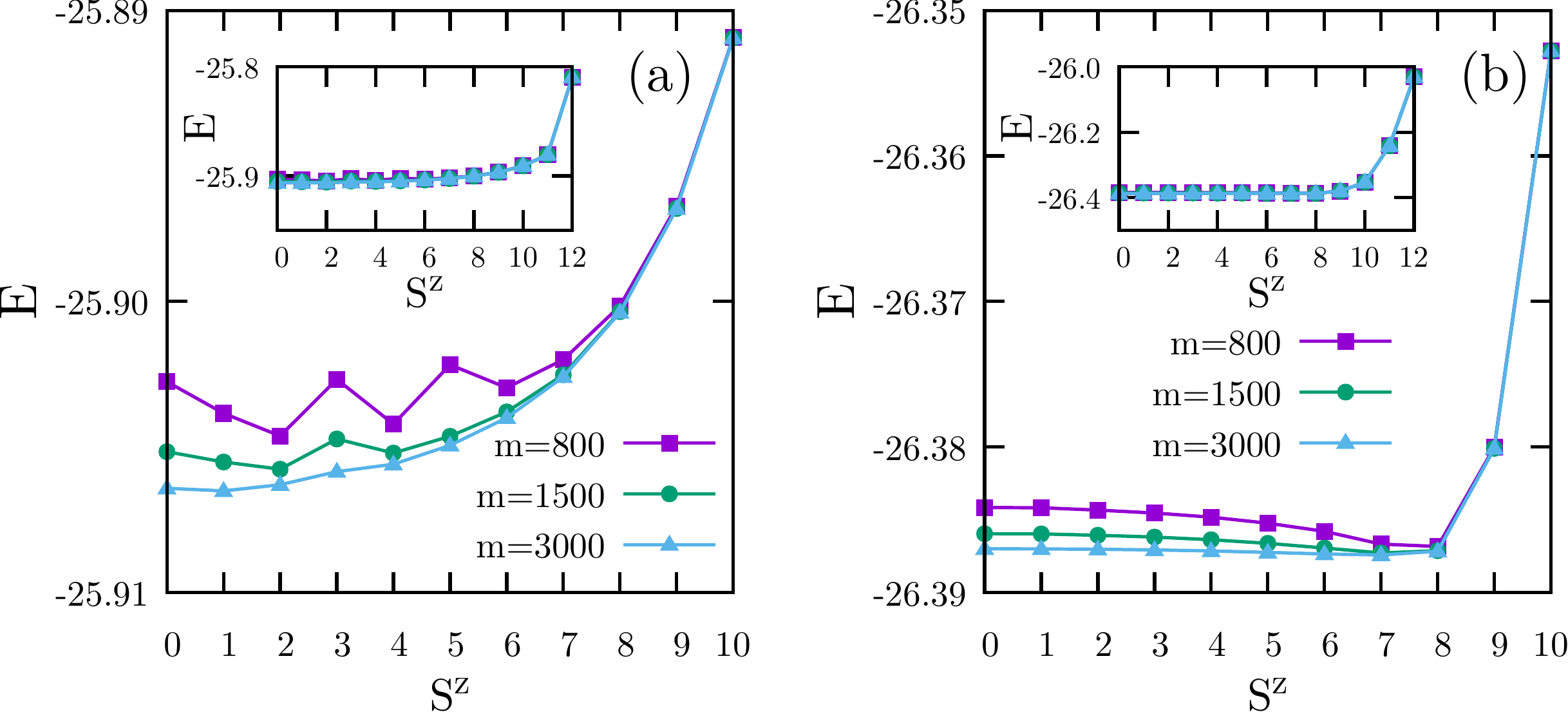}
\caption{Energy ground state as a function of $S_z$ for different number of states $m$ kept in DMRG: (a) for $J' = 0.12 < J'_c$ in the PD phase, (b) for $J' = 0.25 > J'_c$ in the ferrimagnetic phase. The cluster size is $N_s = 12 \times 6$. }
\label{figS2}
\end{center}
\end{figure}

The left and right panels of Fig. \ref{figS2} show the dependence of the ground state energy with the quantum number $S_z$ for the partially disordered phase ($J'= 0.12 < J'_c$) and for the ferrimagnetic phase ($J'= 0.25 > J'_c$), respectively.
In order to illustrate a technical point, in Fig. \ref{figS2}, we display the ground state energy for different $m$ number of states kept in the DMRG computations. 

For $J' < J'_c$, in the PD phase, the ground state has $S_z = 0,$ corresponding to a singlet state. 
There are nearly quasi-degenerate states, corresponding to small $S_z$  values; however, there is an appreciable 
energy difference (of the order of $J'$) between the ground state and the fully saturated ferromagnetic C subsystem, as it is illustrated in the inset of Fig. \ref{figS2} (a).

Within the ferrimagnetic phase $J' > J'_c$ (right panel), there is a degeneracy among states with different $S_z$, up to a maximum $S_z^{\rm max}=8$ . For smaller $m$,  
it seems that the ground state has a $S_z$ quantum number different from zero, however, due to the $SU(2)$ symmetry, this cannot be the case. This non-monotonic behavior of the ground state energy as a function of $S_z$ is related to the fact that for larger $S_z$ the Hilbert space is smaller, and DMRG performs better, yielding a lower ground state energy. Note that, as we increase the number of states $m$, the ground state energy clearly shows a degeneracy for $S_z$ between 0 and $S_z^{\rm max}$, signaling the presence of a ferrimagnetic ground state.  In this way, we determine the value of $S_z^{\rm max}$ for different $J'$.

At $J' = J'_c$, $S_z^{\rm max}$ jumps from zero to a finite value, lower than that corresponding to a fully polarized $C$ subsystem, $\pm \frac{1}{2} N_s/3$. This can be explained considering that, although the local magnetization at the center of the hexagons is very large (see Fig. 3 of the main text), however, it never reaches the classical value 1/2 ($m_C < 0.45$). So, even at the onset of the magnetic ordering of the C spins, quantum fluctuations prohibit that $S_z^{\rm max}$ has its maximum possible value. From Eq. (3) of the main text, we can see that close to $J'_c,$ where
$\theta \simeq \pi$, $S_z^{\rm max}$ has a value of the order of  $m_C N_s/3$. 

\subsection{Stability of the partial disordered phase under a magnetic field}

\begin{figure}[h]
\begin{center}
\includegraphics*[width=0.70\textwidth]{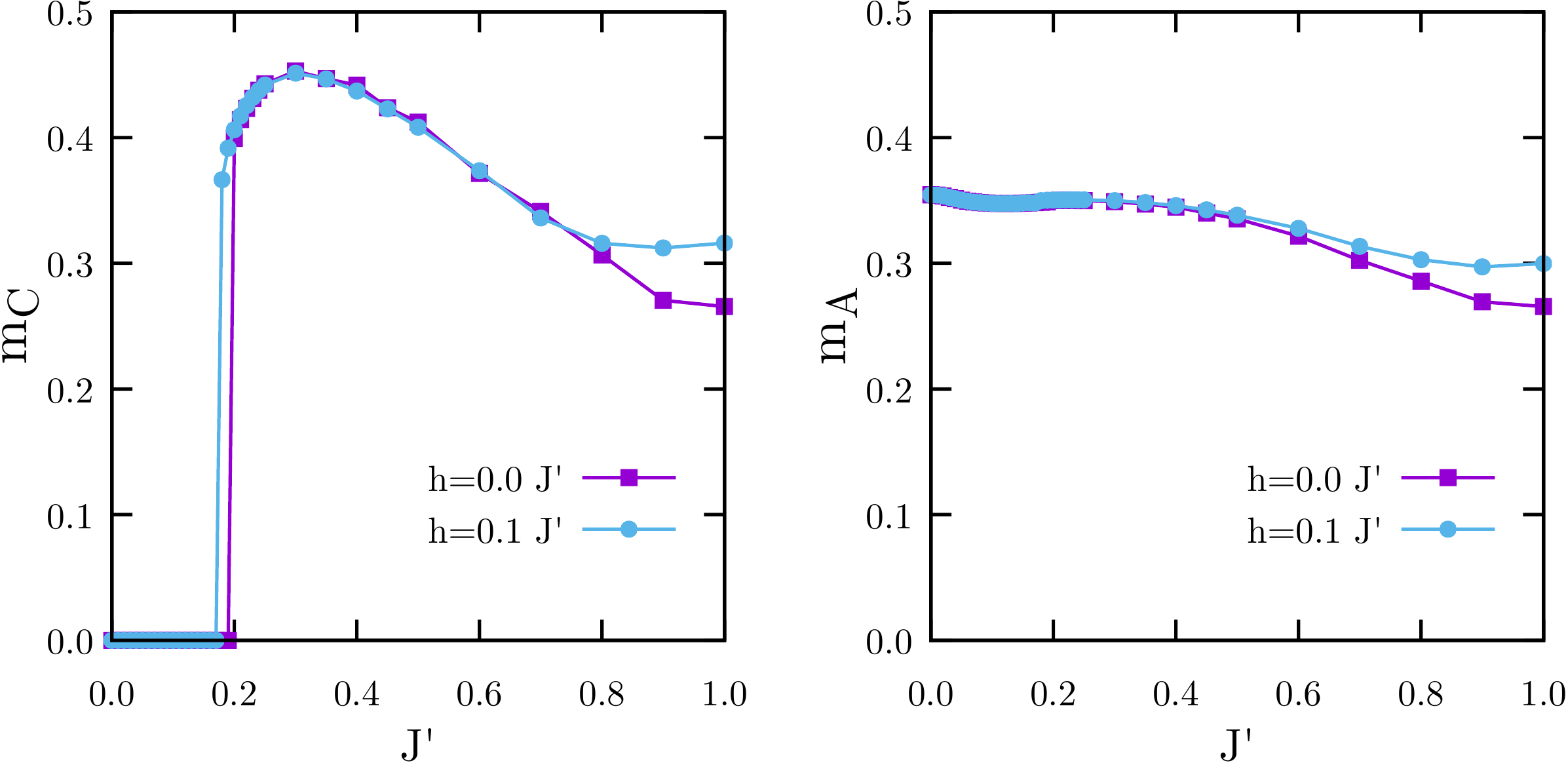}
\caption{Local magnetization of the central spin sublattice $m_C$ (left panel) 
and the honeycomb sublattice $m_A (=m_B)$ (right panel) for the $12 \times 6$ triangular cluster, with and without the application of a uniform magnetic field $h=J'/10$ in the $C$ sublattice only.}
\label{figS3}
\end{center}
\end{figure}

In the above subsection, we have found that the ground state of the partially disordered phase has $S_z  = 0$, with other $S_z$ states close in energy. In order to check the robustness of the vanishing of the $m_C$ local magnetization in the PD phase, we have applied a uniform small magnetic field $h=J'/10$ only to the $C$ central spins. In this way, we would favor any tendency of the central spins to align ferromagnetically, while the honeycomb sublattice would remain unperturbated. However, as it is shown in the left panel of Fig. \ref{figS3}, $m_C = 0$ even in the presence of the magnetic field $h$, 
departing its behavior from the zero-field case only for $J'$ very close to $J'_c$. On the other hand, $m_A (=m_B)$ (right panel) does not change at all in the PD phase. Note that in the previous known partially disordered phases (present in certain Ising and XY-like models), the orphan spins are uncorrelated and, as a consequence, the application of a small magnetic field will be enough to order them ferromagnetically. In our model, the correlated singlet nature of the PD phase makes the system robust against the magnetic field perturbation.

Close to the isotropic point ($J'=J$), the effect of the magnetic field is appreciable because $h$ becomes an important fraction of $J,$ reducing quantum fluctuations in both magnetic sublattices, and, 
as a consequence, the local magnetizations $m_A$ and $m_C$ increase with $h$.

\subsection{Convergence with the number of states kept and finite size effects}

\begin{figure}[h]
\begin{center}
\includegraphics*[width=0.70\textwidth]{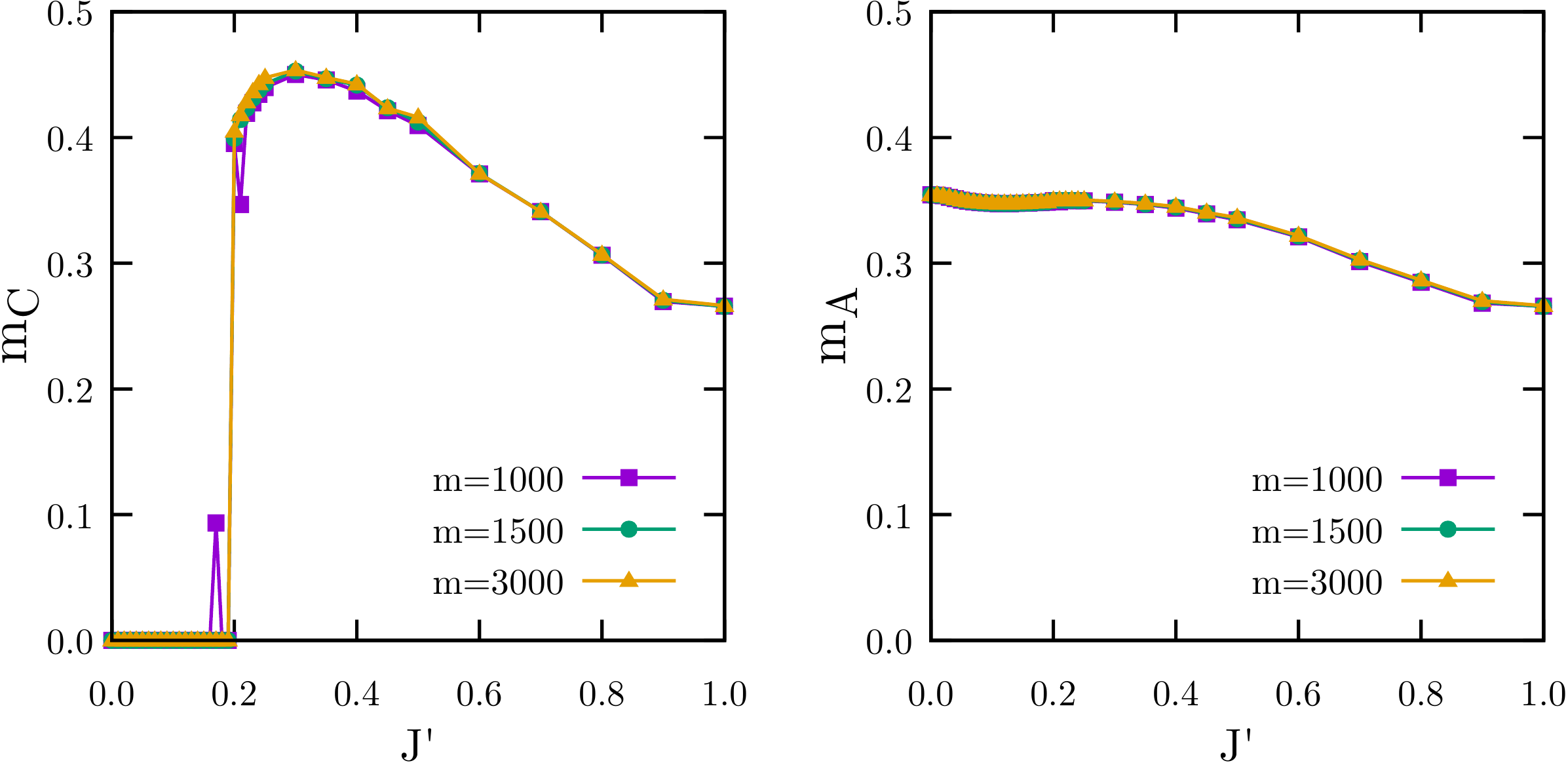}
\caption{Local magnetization $m_C$ (left) and $m_A (=m_B)$ (right) as a function of $J'$ for the triangular cluster $12 \times 6$, for different number $m$ of kept states. }
\label{figS4}
\end{center}
\end{figure}

To analyze the convergence of the DMRG predictions with the number $m$ of states kept, we show in Fig. \ref{figS4} 
the behavior of the local magnetizations for different $m$, using the cluster with $L_x=12 \times L_y=6$ sites. 
It can be seen an excellent quantitative agreement for the three values of $m$ presented, for almost all $J'$. Only for $m=1000$, $m_C$ has some noisy behavior close to $J'_c$, that dissappears for larger $m$. 
It is worth to emphasize that the numerical results in our model converge for a relatively small value of $m$ in comparison to what happens in highly frustrated 
antiferromagnets, like the kagom\'e Heisenberg model. This fact would signal that in the PD phase for $J'\neq 0$ there is a low effective degeneracy, albeit the system has an extensive degeneracy for $J'=0$.

\begin{figure}[h]
\begin{center}
\includegraphics*[width=0.70\textwidth]{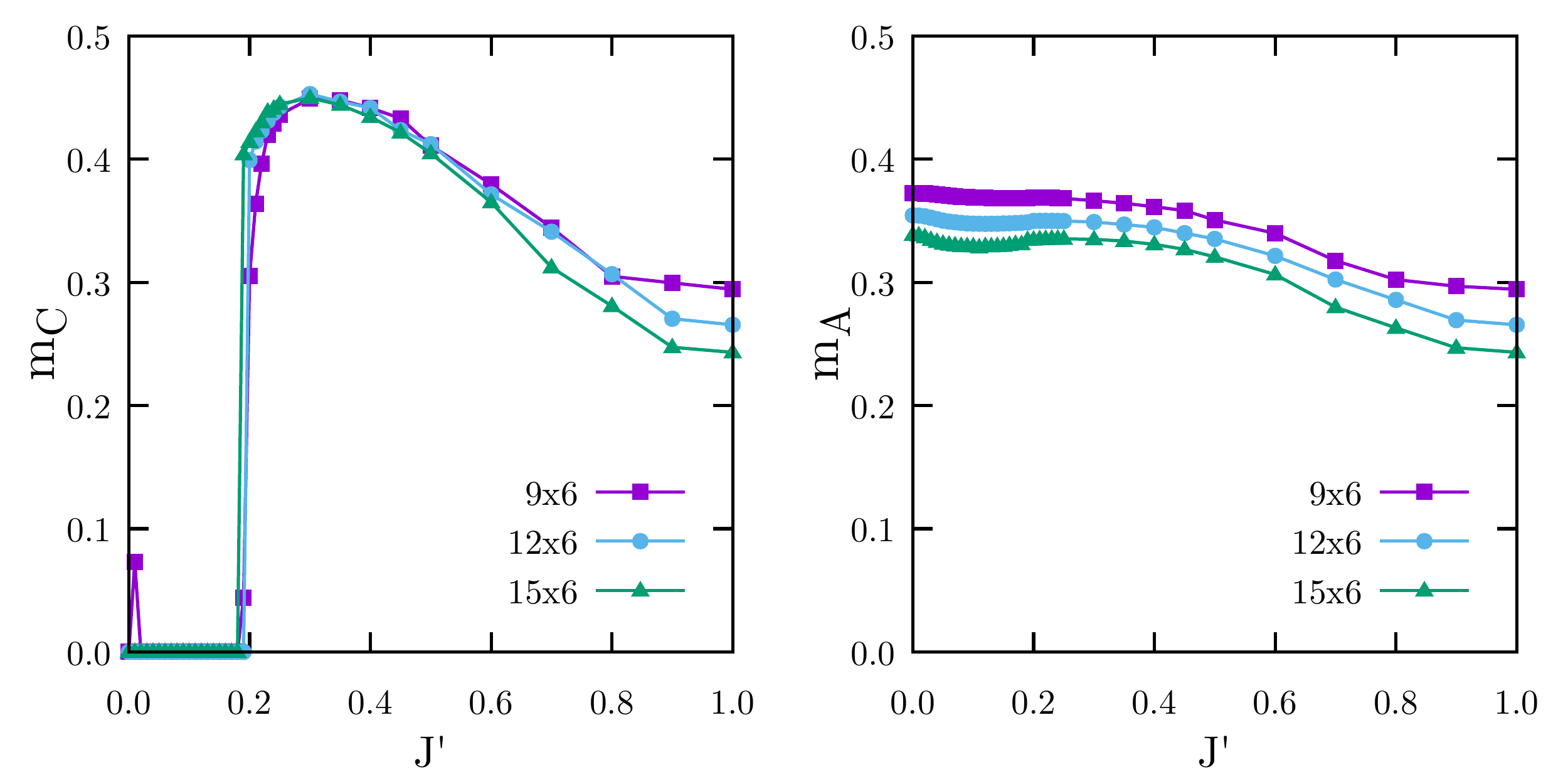}
\caption{Local magnetization $m_C$ (left) and $m_A (=m_B)$ (right) as a function of $J'$ for triangular clusters with $L_y=6$ and different $L_x$.}
\label{figS5}
\end{center}
\end{figure}

Once we are confident that for $m \gtrsim 1500$ the DMRG predictions are well converged, we focus on the finite size effects. 
We present in Fig.~\ref{figS5} the local magnetizations $m_A$ and $m_C$ as a function of  $J'$ for clusters with $L_y=6$ legs and different 
$L_x$, using $m=1500$ states. Clearly, the finite size effects are, in general, not negligible, although the overall behavior of both 
magnetizations does not change with $L_x$. Most relevant for our findings is the fact that, for $ 0 \le J' \lesssim 0.5$, $m_C$ exhibits 
rather small finite size effects, resulting,  as a consequence, that the partially disordered phase survives in the thermodynamic limit, 
with a reliable critical value $J'_c \simeq 0.18$. 
On the other hand, the honeycomb sublattice has 
appreciable finite size effects for all $J'$, the same happens for the central spins close to the isotropic point $J'=J$. In these cases, an 
appropriate thermodynamic limit extrapolation \cite{white07b} would yield reduced, but finite, local magnetizations ($m_A=m_C \simeq 0.205$ for the isotropic
point \cite{white07b}, $m_A \simeq 0.268$ for the honeycomb lattice). 

\begin{figure}[h]
\begin{center}
\includegraphics*[width=0.35\textwidth]{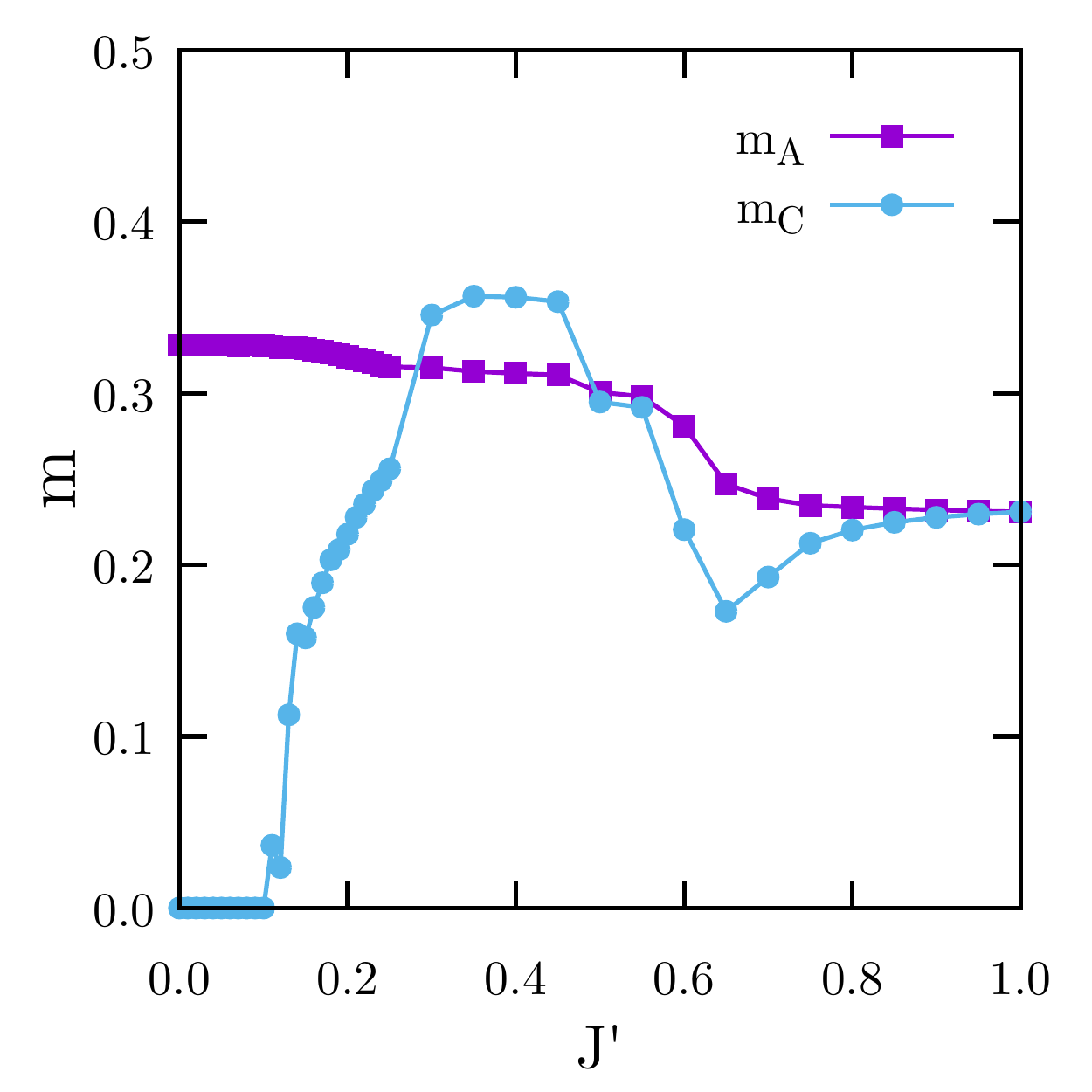}
\caption{Local magnetization of the honeycomb sublattice $m_A (=m_B)$ and of the C sublattice $m_C$ 
for the $12 \times 4$ triangular cluster.}
\label{figS6}
\end{center}
\end{figure}

With respect to the stability of our results for clusters with different $L_y$, in Fig. \ref{figS6} we present the local magnetizations for the $L_x=12$, $L_y=4$ triangular cluster. It can be seen that for this smaller cluster, there is also a transition to a partially disordered phase 
below $J'_c \simeq 0.18$ and there are similar behaviors of the local magnetization of both subsystems.
We have checked that the $C$ spins are also ferromagnetically correlated 
(with an almost constant correlation below $J'_c$) for this cluster. 
Besides, we have found that close to the isotropic triangular point  ($J'  \simeq 0.7$), the magnetization curve for the $C$ subsystem has a shallow dip, probably due to the quasi-one-dimensional character of this cluster (as it was discussed for the isotropic triangular Heisenberg model \cite{white07b}). 

\begin{figure}[h]
\begin{center}
\includegraphics*[width=0.70\textwidth]{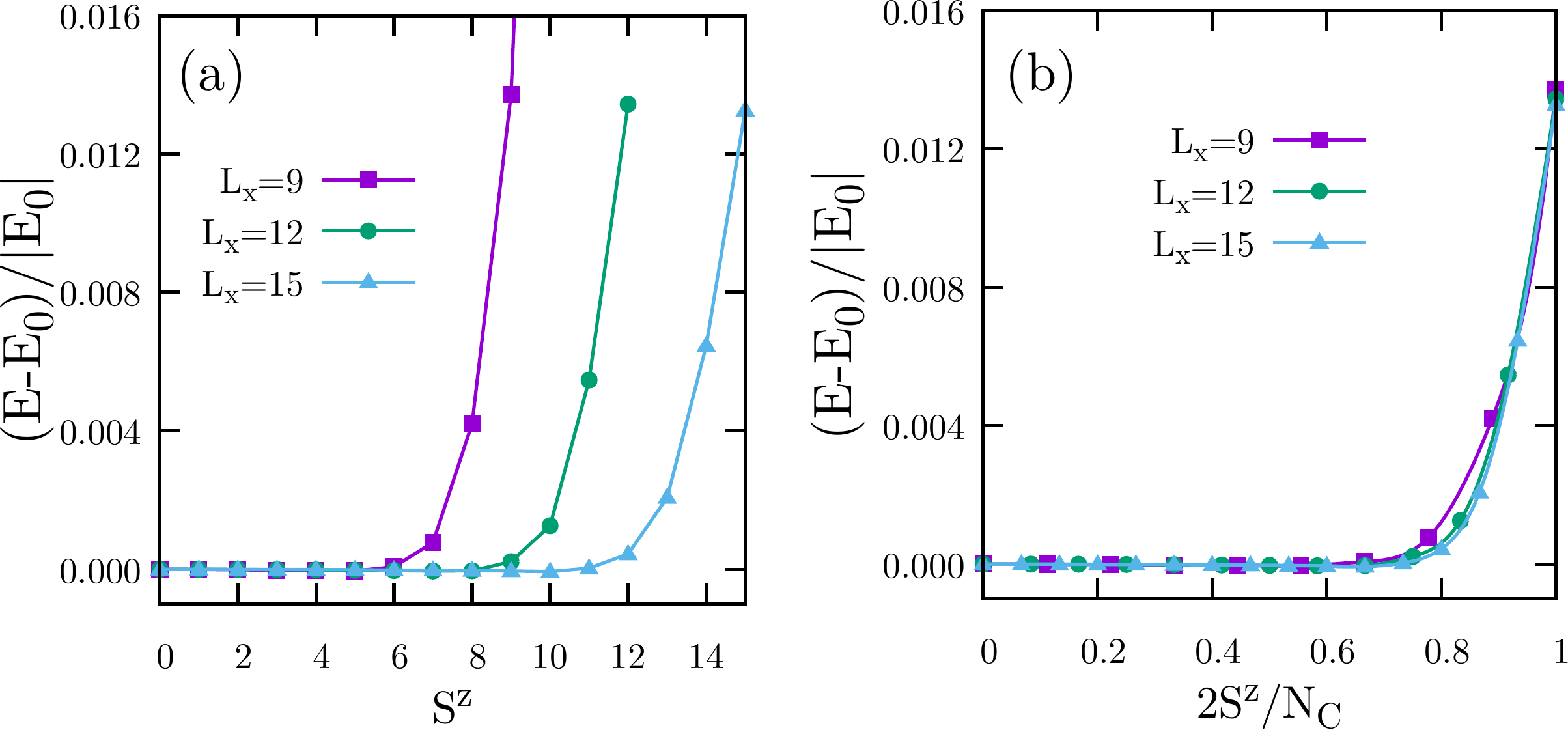}
\caption{DMRG ground state energy as function of (a) the quantum number $S_z$  and (b) the quantum number $S_z$ normalized by the number $N_C$ of C spins, for different cluster sizes with $L_y=6$.}
\label{figS7}
\end{center}
\end{figure}

Regarding the finite size effects in the ferrimagnetic state, Figs. \ref{figS7} 
display the behavior of the ground state energy for $J'=0.25$  as a function of $S_z$ for cluster sizes corresponding to $L_y=6$ and different $L_x$'s.  
If we normalize $S_z$ with the number of $C$ sites, $N_C$ (right panel of Fig. \ref{figS7}), 
the value of the normalized $S_z^{\rm max}$ is almost independent of $L_x$. 
This gives us confidence about the relevance of our results in the thermodynamic limit.

\subsection{Spin-spin correlations between the magnetic subsystems in the partially disordered phase}

\begin{figure}[ht]
\begin{center}
\includegraphics*[width=0.35\textwidth]{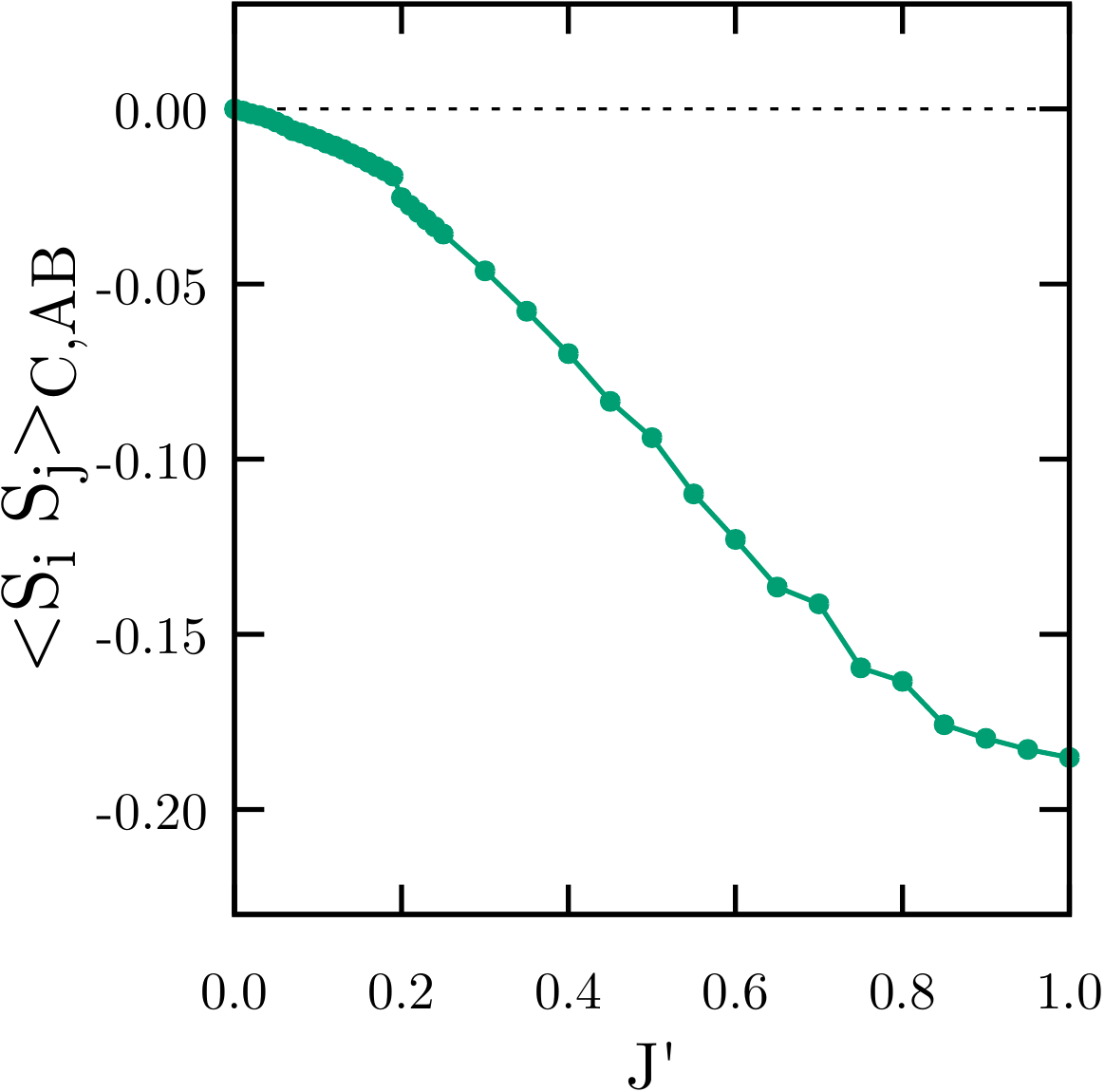}
\caption{Nearest neighbor spin-spin correlation between the honeycomb subsystem $A (B)$ spins and the center of hexagon $C$ spins.}
\label{figS8}
\end{center}
\end{figure}

Fig. \ref{figS8} shows the (averaged) nearest neighbor spin-spin correlation between the C spins and the A (or B) honeycomb spins as a function of $J'$. It can be seen that, except for $J'=0$, there exists a finite antiferromagnetic correlation between the two subsystems, even in the partially disordered phase, $ 0 \le J' \le 0.18$. This result reinforces the correlated character of the partial disorder phase, that is, the orphan spins do not behave as a perfect paramagnet like in previous known PD phases. Furthermore, the finite correlation between subsystems supports our arguments about the Casimir-like effect that C spins are ferromagnetically aligned at short distance \emph{via} the zero-point quantum fluctuations of the magnetically ordered honeycomb subsystem. 

\begin{thebibliography}{99}
\bibitem{magnet} As we are working on finite size clusters, the local magnetizations are somewhat higher than 
the expected ones in the thermodynamic limit: $m \simeq 0.20$ for the triangular lattice \cite{capriotti99,white07}, 
and $m \simeq 0.26$ for the honeycomb case \cite{oitmaa92}.
\bibitem{average} We compute this correlation averaging over all the first neighbors of the cluster, 
except those closest to the open edges in order to avoid border effects.
\bibitem{campo} Another evidence of the absence of long-range ferromagnetic order in the central spin subsystem, below $J'_c$, is given by its zero magnetization even under the application of a small uniform magnetic field $h$ to this subsystem. See \cite{sm}.
\end{thebibliography}
\end{document}